\documentclass[12pt]{article}
\usepackage{graphicx}
\usepackage{epsfig}
\usepackage{latexsym}
\usepackage{latexsym}
\usepackage{amssymb}
\usepackage{amsmath}
\newcommand{\labell}[1]{\label{#1}}
\newcommand{\reef}[1]{(\ref{#1})}

\def\ie{{\it i.e.}}

\DeclareSymbolFont{AMSb}{U}{msb}{m}{n}
\DeclareMathSymbol{\IN}{\mathbin}{AMSb}{"4E}
\DeclareMathSymbol{\IZ}{\mathbin}{AMSb}{"5A}
\DeclareMathSymbol{\IR}{\mathbin}{AMSb}{"52}
\DeclareMathSymbol{\Q}{\mathbin}{AMSb}{"51}
\DeclareMathSymbol{\II}{\mathbin}{AMSb}{"49}
\DeclareMathSymbol{\IC}{\mathbin}{AMSb}{"43}
\DeclareMathSymbol{\IP}{\mathbin}{AMSb}{"50}
\DeclareMathSymbol{\IH}{\mathbin}{AMSb}{"48}
\DeclareMathSymbol\IA{\mathalpha}{AMSb}{"41}
\DeclareMathSymbol\IS{\mathalpha}{AMSb}{"53}

\def\Q{{\cal Q}}

\begin{document}

\begin{flushright}
{\tt hep-th/0610223}
\end{flushright}
{\flushleft\vskip-1.35cm\vbox{\psfig{figure=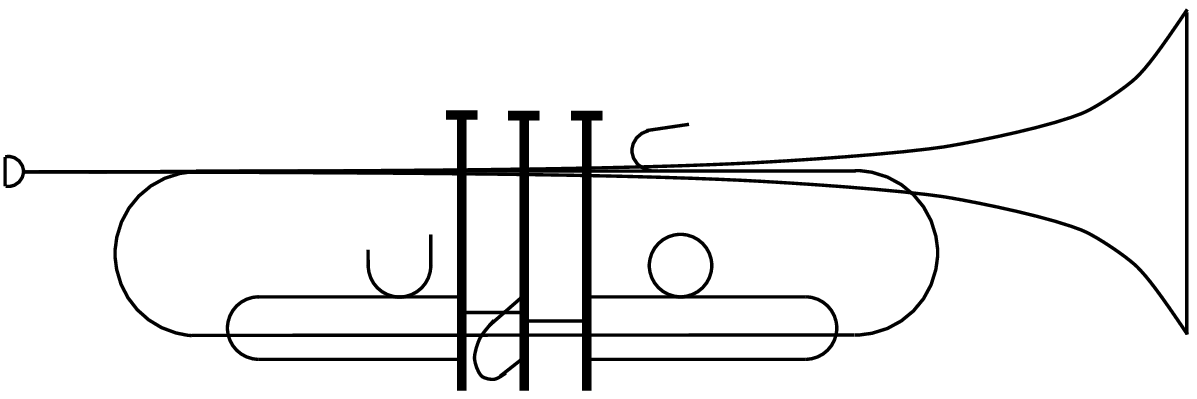,height=0.45in}}}
\bigskip
\bigskip
\bigskip
\bigskip

\begin{center} {\Large \bf String Theory Without Branes}





\end{center}

\bigskip \bigskip \bigskip

\centerline{\bf Clifford V. Johnson}

\bigskip
\bigskip

  \centerline{\it Department of Physics and Astronomy }
\centerline{\it University of
Southern California}
\centerline{\it Los Angeles, CA 90089-0484, U.S.A.}

\bigskip

\centerline{\small \tt  ${}^1$johnson1 [at] usc.edu}

\bigskip
\bigskip


\begin{abstract}
  
  We present a class of solvable models that resemble string theories
  in many respects but have a strikingly different non--perturbative
  sector. In particular, there are no exponentially small
  contributions to perturbation theory in the string coupling, which
  normally are associated with branes and related objects.
  Perturbation theory is no longer an asymptotic expansion, and so can
  be completely re--summed to yield all the non--perturbative physics. We
  examine a number of other properties of the theories, for example
  constructing and examining the physics of loop operators, which can
  be computed exactly, and gain considerable understanding of the
  difference between these new theories and the more familiar ones,
  including the possibility of how to interpolate between the two
  types.  Interestingly, the models we exhibit contain a family of
  zeros of the partition function which suggest a novel phase
  structure. The theories are defined naturally by starting with
  models that yield well--understood string theories and allowing a
  flux--like background parameter to take half--integer rather than
  integer values. The family of models thus obtained are seeded by
  functions that are intimately related to the classic rational
  solutions of the Painlev\'e~II equation, and a family of
  generalisations.

\end{abstract}
\newpage \baselineskip=18pt \setcounter{footnote}{0}

\section{Background}
\label{sec:introduction}

With a few notable (and highly instructive) examples in $D\leq2$,
string theory still lacks a satisfactory and well--understood
non--perturbative definition.  It is fair to say that while strings
have marvellous properties that may prove a great boon for studying
Nature, we have not been learning about these properties
systematically, but instead by following the theory into various
regimes which have become accessible to us by various techniques. As a
result, it is not clear what the big picture is ---certainly not clear
is the complete list of phenomena we should expect from string theory.

Among the phenomena which would now be firmly on everybody's list
---especially in the context of non--perturbative physics--- are
branes of various types, manifesting themselves as dynamical extended
objects, and leaving a characteristic signature in small $g_s$
perturbation theory as exponentially small corrections,
$\exp(-C/g_s^p)$ (where $C$ is $g_s$ independent and $p=1,2$).
 The argument of the exponent is the action of the
brane. Most ubiquitous these days are of course the D--branes ($p=1$), which
by virtue of their ease of handling in terms of open string boundary
conditions, have become a rather standard feature in discussions of
non--perturbative string theory.

As is well known, some of the earliest signs of D--branes' role in
non--perturbative string physics was in the aforementioned context of
$D\leq2$ theories\cite{Shenker:1990uf}. In fact, an examination of the
literature might lead the reader to conclude that there {\sl must} be
$\exp(-C/g_s^p)$ effects in any string theory. According to the
argument of ref.\cite{Shenker:1990uf}, these effects are tied to the
characteristic rapid growth of string perturbation theory at large
orders in the expansion about small $g_s$.

In the light of this, one may ask the simple question: Is it possible
that one can have a string theory {\it without} these features? What
kind of properties would the theory have to possess, and if it has
those properties, does it still qualify to be considered as a string
theory? The results presented in this paper provoke the question, and
presents several models as primary exhibits to be considered.  This
will all be in the context of highly solvable models, so, having
answered those questions, a final one would be: Might there be such
phenomena in the context of other and more ``realistic'' string
theories that are worth looking out for? We end this paper by trying
to rephrase our finding in a reasonably model--independent manner, in
order to help answer this important broader question.

\section{The Models}
\subsection{The Framework}
A class of well understood string theories can be given a rather
thorough path integral definition by regulating the sum over world
sheet metrics with a discretisation supplied by a 
model of complex
matrices\cite{Dalley:1992qg,Dalley:1992vr,Dalley:1992yi,Dalley:1992br}.
(See also ref.\cite{Johnson:2004ut} for a collection of some of the
results.) The continuum limit, the all-genus expansion, and the
physics beyond the genus expansion can all be extracted by taking the
size of the matrices to infinity in the vicinity of critical points of
the model that yield the universal physics, in the spirit of the
original double scaling limit of
refs.\cite{Brezin:1990rb,Douglas:1990ve,Gross:1990vs}. The reader
needn't know the details of the procedure to proceed. In suffices to
know that the output can be cast succinctly in terms of an elegant
family of differential equations, together with some other structures
that we will bring out later as we need them.

Let us start with the  differential equation:
\cite{Dalley:1992qg,Dalley:1992br}:
\begin{equation}
u{\cal R}^2-\frac{1}{2}{\cal R}{\cal R}^{''}+\frac{1}{4}({\cal
R}^{'})^2
  =\nu^2\Gamma^2\ .\labell{eq:nonpert}
\end{equation}
where $u(z)$ is a real function of the real variable $z$; a prime
denotes $\nu \partial/\partial z$; and $\Gamma$ and $\nu$ are
constants. The quantity $\mathcal{R}$ is defined by:
\begin{equation}
\labell{eq:R} {\cal
  R}=\sum_{k=0}^\infty  \left(k+\frac{1}{2}\right)t_k R_k[u]\ ,
\end{equation}
where the $R_k[u]$ ($k=0,\ldots$) are polynomials in $u(z)$ and its
$z$--derivatives, called the Gel'fand--Dikii polynomials. They are
related by a recursion relation\cite{Gelfand:1975rn}:
\begin{equation}
  \labell{eq:recursion}
  R^{'}_{k+1}=\frac{1}{4}R^{'''}_k-uR^{'}_k-\frac{1}{2}u^{\prime}R_k\ ,
\end{equation}
and are fixed by the constant $R_0$, and the requirement that the rest
vanish for vanishing $u$. Some of them are:
\begin{equation}
  R_0=\frac{1}{2}\ ;\quad R_1=-\frac{1}{4}u\ ;\quad R_2=\frac{1}{16}(3u^2-u^{''})\ ;\quad R_3=-\frac{1}{64}\left(10u^3- 10u u^{''}-5(u^{'})^2+u^{''''}\right)\ ;\hdots   \labell{eq:firstfew}
\end{equation}
The $k$th model is chosen by setting all the other $t \,$s to zero
except $t_0\equiv z$, and $t_k$, the latter being fixed to a numerical
value  such that in ${\cal R}=c{\cal R}_k-z$, $c$ sets the coefficient of
$u^k$ to unity.

\subsection{Old Choices}
\label{sec:oldchoices}

The function $u(z)$ defines the partition function $Z=\exp(-F)$ of the
string theory {\it via}:
\begin{equation}
u(z)=2\nu^2\frac{\partial^2 F}{\partial \mu^2}\Biggl|_{\mu=z}\ ,
  \labell{eq:partfun}
\end{equation}
where $\mu$ is the coefficient of an operator in
the world--sheet theory.  

From the point of view of the $k$th theory, the other $t_j$s
represent coupling to closed string operators ${\cal O}_j$. It is
well known that the insertion of each operator can be expressed in
terms of the KdV flows\cite{Douglas:1990dd,Banks:1990df}:
\begin{equation}
  \labell{eq:kdvflows}
  \frac{\partial u}{\partial t_j}= R^{'}_{j+1}\ .
\end{equation}
The operator ${\cal O}_0$ couples to $t_0$, which is in fact $-4z$.
So $u(z)$ is the two--point function of ${\cal O}_{0}$.

For the $k$th model, equation~\reef{eq:nonpert}, which has remarkable
properties\cite{Carlisle:2005wa,Carlisle:2005mk}, is known to supply a
complete non--perturbative definition of a family of spacetime bosonic
string
theories\cite{Dalley:1992qg,Dalley:1992vr,Dalley:1992yi,Dalley:1992br,Johnson:1992pu}.
The models are actually type~0A strings\cite{Klebanov:2003wg}, based
upon the $(4k,2)$ superconformal minimal models coupled to
superliouville theory. The relevant solutions have asymptotics:
\begin{eqnarray}
  u(z)&=&z^{\frac{1}{k}}+\frac{\nu\Gamma}{kz^{1-\frac{1}{2k}}}+\cdots\quad \mbox{\rm for}\quad z\longrightarrow +\infty\ ,
\nonumber\\
u(z)&=&\frac{\nu^2\left(\Gamma^2-\frac14\right)}{z^2}+\cdots\quad \mbox{\rm for}\quad z\longrightarrow -\infty\ .
  \labell{eq:largez}
\end{eqnarray}
Integrating twice, the asymptotic expansions in
equations~\reef{eq:largez} furnish the free energy perturbatively as
an expansion in the dimensionless string coupling:
\begin{equation}
  \labell{eq:stringcoupling}
g_s=\frac{\nu}{\mu^{1+\frac{1}{2k}}}\ .
\end{equation}

For these models, in the $\mu\to+\infty$ regime, $\Gamma$ represents
the number of background ZZ D--branes\cite{Zamolodchikov:2001ah} in
the model, with a factor of $\Gamma$ for each boundary in the
worldsheet expansion. These are point--like branes that are localized at
infinity far in the strong coupling regime in the Liouville directions
$\varphi$. In the $\mu\to-\infty$ regime, $\Gamma$ represents the
number of units of R--R flux in the background, with $\Gamma^2$
appearing when there is an insertion of pure R--R
flux\cite{Dalley:1992br,Klebanov:2003wg}.

Since there is a unique non--perturbative solution connecting the two
regimes, the equation supplies a non--perturbative completion of the
theory that is an example of a geometric
transition\cite{Gopakumar:1998ki} between these two distinct
(D--branes {\it vs.} fluxes) spacetime descriptions of the physics.

That $\Gamma$ is integer and positive in these models (matching the
physics interpretation above) is naturally encoded in the properties
of the equation. (See refs.\cite{Carlisle:2005wa,Carlisle:2005mk} for
more on this.) The equations' solutions, with the asymptotics given
above, are pole--free for integer $\Gamma$.

\subsection{B\"{a}cklund Transformation}
An attractive feature of the system is that the solutions of the
equation for $\Gamma$ differing by an integer are connected by the
relation\cite{Carlisle:2005wa,Carlisle:2005mk}:
\begin{eqnarray} \labell{eq:Back-Explicit-Gam}
u_{\Gamma \pm 1} = \frac{3 \left(\mathcal{R}^{\prime} \right)^2 -
2 \mathcal{R}\mathcal{R}^{\prime \prime} \mp 8 \nu\Gamma \,
\mathcal{R}^{\prime} + 4 \nu^2\Gamma^2}{4 \mathcal{R}^2}\ ,
\end{eqnarray}
where $\mathcal{R} \equiv \mathcal{R}(u_{\Gamma})$, and from this,
starting with $u_{\Gamma=0}$ it is easy to see that
$u_{\Gamma=+1}=u_{\Gamma=-1}$ and by extension
$u_{\Gamma}=u_{-\Gamma}$, which is a statement of the charge
conjugation invariance of the theory. In fact, the
equation~\reef{eq:Back-Explicit-Gam} defines the celebrated B\"acklund
transformations of the underlying KdV hierarchy of equations~\reef{eq:kdvflows}.

Before we proceed further, let us note an additional set of facts that
will be pertinent. The following ``Muira'' transformation\cite{Carlisle:2005mk}
\begin{eqnarray}
u_\Gamma&=&v_C^2-v_C^\prime= v_{1-C}^2-v_{1-C}^\prime\ ,
\end{eqnarray}
defines a function $v_C$, or a function ${\bar v}_C\equiv v_{1-C}$, a
solution of the equations defining the Painlev\'{e}~II hierarchy. This
hierarchy can be written, if we define the quantities $S_k$ this way
\begin{eqnarray} \labell{eq:DJMW-Poly}
S_k \equiv \frac12 R_k^\prime [v^2 - v^\prime] - v R_k[v^2 - v^\prime]\ ,
\end{eqnarray}
as
\begin{equation} \labell{eq:DJMW-5}
\sum_{k=1}^{\infty} \left(k + \frac12\right) t_k S_k [v(z)] + z v(z) = \nu C\ ,
\end{equation}
where $C=\frac{1}{2}\pm\Gamma$, and $v(z)$ is $v_C$ or ${\bar
  v}_C\equiv v_{1-C}$. The B\"{a}cklund transformations on $u(z)$ are
accompanied\cite{Carlisle:2005mk} by an explicit set on $v(z)$ (here
we use $v_\Gamma$ for $v_C$ and ${\bar v}_\Gamma$ for ${\bar v}_C$):
\begin{eqnarray} \labell{eq:vBackExplicit}
v_{\Gamma-1} = - v_\Gamma + \frac{2 \nu \Gamma}{\mathcal{R}[v^2_\Gamma - v^{\prime}_{\Gamma}]} \, , \quad
\bar{v}_{\Gamma-1} = - \bar{v}_\Gamma - \frac{2 \nu (\Gamma - 1)}{\mathcal{R}[\bar{v}^2_\Gamma + \bar{v}^{\prime}_{\Gamma}]} \nonumber \\
v_{\Gamma+1} = - v_\Gamma + \frac{2 \nu (\Gamma + 1)}{\mathcal{R}[v^2_\Gamma + v^{\prime}_{\Gamma}]} \, , \quad
\bar{v}_{\Gamma+1} = - \bar{v}_\Gamma - \frac{2 \nu \Gamma}{\mathcal{R}[\bar{v}^2_\Gamma - \bar{v}^{\prime}_{\Gamma}]}
\end{eqnarray}

The function $v(z)$ plays a natural role in the problem, furnishing a
solution to the $\lambda=0$ sector of the eigenvalue problem
\begin{eqnarray}
\labell{eq:eigenvalue}
{\cal H}\psi=\left( -\nu^2\partial^2 +u(z)\right)\psi=\lambda \psi\ ,
\end{eqnarray}
which is 
\begin{equation}
\psi= \exp\left\{ -\frac{1}{\nu}\int v(z) dz\right\} = e^{-{\cal F}}\ .
\end{equation}
The wavefunction $\psi$ is in fact\cite{Klebanov:2003wg} the partition function of a FZZT
D--brane, which lies along the Liouville direction, starting at
$\varphi=-\infty$ and stretching to $\varphi=-\ln\lambda$, where it
terminates. The $\lambda=0$ case represents the tip of the brane
stretching all the way to the $\Gamma$ ZZ D--branes at
$\varphi=+\infty$.

\subsection{New Choices}

Now let us explore something new, emboldened by the successful manner
in which the properties of the string equation and its solutions
encode so neatly the physics of the type~0A string, with little need to
refer to the underlying matrix model which originally produced the equation.

It is rather curious that the case of $\Gamma$ being half integer
corresponds to a special and well known case, on the Painlev\'e II
side of things. We'll focus on the case of $k=1$ for a while,
whence the equation for $v(z)$ is indeed the classic Painlev\' e~II
equation:
\begin{equation}
  \labell{eq:painleveII}
  \frac12 v^{\prime\prime}-v^3+zv+\nu\left(\frac12-\Gamma\right)=0 .
\end{equation}
When $\Gamma$ is half--integer, it is well known that there are {\it
  exact} rational solutions to this equation\cite{airault}. The
simplest case is $\Gamma=\frac12$ for which an obvious solution is
$v=0$. Then $u=v^2-v^\prime=0$ and so the free energy vanishes, and
the partition function is unity. There's not much to see here.

The next simplest case is $\Gamma=\frac32$, and the exact solution can
be determined by eye to be $v(z)=\nu/z$. From this we get:
\begin{equation}
  \labell{eq:simplecasea}
  u_{\Gamma=\frac32}=v^2-v^\prime=\frac{2\nu^2}{z^2}\ ,
\end{equation} and we can check that:
\begin{equation}
  \labell{eq:simplecaseb}
  u_{\Gamma=\frac12}=v^2+v^\prime=0\ .
\end{equation}
This case is also rather simple, but interesting, since the free
energy and partition function are exactly:
\begin{eqnarray}
  \labell{eq:freeparti}
  F&=&-\ln|\mu|\nonumber\\
Z&=&|\mu|\ .
\end{eqnarray}
This simplicity hides a lot of interesting structure that we'll uncover
shortly, especially when we study examples of other $k$. For now, let
us examine the next few $k=1$ cases.

The case of $\Gamma=\frac52$ can be generated using our B\"acklund
transformations for either $u$ or $v$. They give:
\begin{eqnarray}
  \labell{eq:nextcase52}
  v(z)&=&-\frac{2\nu(z^3+\nu^2)}{z(z^3-2\nu^2)}\nonumber\\
u(z)&=&\frac{6z\nu^2(z^3+4\nu^2)}{(z^3-2\nu^2)^2}\ .
\end{eqnarray}
It turns out that we can find the free energy and partition function exactly:
\begin{eqnarray}
  \labell{eq:freeparti52}
  F&=&-\ln|\mu^3-2\nu^2|\nonumber\\
Z&=&|\mu^3-2\nu^2|\ .
\end{eqnarray}
It is interesting to perturbatively expand the free energy:
\begin{equation}
  \labell{eq:expandfree52}
  F=-3\ln|\mu|+2g_s^2+2g_s^4+\frac{8}{3}g_s^6+4g_s^8+\frac{32}{5}g_s^{10}+\cdots
\end{equation}
and see that it has a worldsheet expansion which is entirely in terms
of closed strings. We notice that there are no open strings
(background D--branes) in the expansion and the fact that the
expansion can be re--summed exactly means that there are no
exponentially small contributions to this expansion. In other words,
there are no D--branes in the theory, as we shall demonstrate further
in more detail in the sequel.

Order $\exp(-1/g_s)$ effects are tied to the rapid growth of
perturbation theory with increasing powers in $g_s$. We therefore
conclude that it does not grow rapidly. In fact, writing the partition
function in terms of the string coupling, we see that it stops at
finite order.

The next example is $\Gamma=\frac72$, for which we find:
\begin{eqnarray}
  \labell{eq:uv72}
  v&=&\frac{3z^2\nu(z^6-4z^3\nu^2+40\nu^4)}{(z^3-2\nu^2)(z^6-10z^3\nu^2-20\nu^4)}\ ,\nonumber\\
u&=&\frac{12z\nu^2(z^9+150z^3\nu^4-200\nu^6)}{(z^6-10z^3\nu^2-20\nu^4)^2}\ .
\end{eqnarray}
These seem to be getting rather messy, but after a bit of thought, the
second function can indeed be integrated to yield:
\begin{eqnarray}
  \labell{eq:freeparti72}
  F&=&-\ln|\mu^6-10\mu^3\nu^2-20\nu^4|\ ,\nonumber\\
  Z&=&|\mu^6-10\mu^3\nu^2-20\nu^4|\ .
\end{eqnarray}
The perturbative expansion of the free energy is:
\begin{equation}
  \labell{eq:expandfree72}
  F=-6\ln|\mu|+10g_s^2+70g_s^4+\frac{1600}{3}g_s^6+4700g_s^8+44000g_s^{10}+\cdots
\end{equation}
As a final example, we have the case of $\Gamma=\frac92$, for which:
\begin{eqnarray}
v&=&\frac{\nu\left(4z^{15}-25z^{12}\nu^2+250z^9\nu^4+2800z^6\nu^6-7000z^3\nu^8+7000\nu^{10}\right)}{z(z^9-30z^6\nu^4-1400\nu^6)(z^6-10z^3\nu^2-20\nu^4)}\ , \nonumber\\
   u&=&\frac{20\nu^2(z^{18}-24 z^{15}\nu^2+630z^{12}\nu^4+9800z^9\nu^6-117600z^6\nu^8+196000\nu^{12})}{(z^9-30z^6\nu^4-1400\nu^6)^2z^2}\ .
 \labell{eq:uv92}
 \end{eqnarray}
Remarkably this all integrates to give:
\begin{eqnarray}
  \labell{eq:freeparti92}
  F&=&-\ln|\mu^{10}-30\mu^7\nu^2-1400\mu\nu^6)|\nonumber\\
&=&-10\ln|\mu|+30g_s^2+450g_s^4+10400g_s^{6}+244500g_{s}^{8}+6120000g_{s}^{10}\cdots\ , \nonumber\\
Z&=&|\mu^{10}-30\mu^7\nu^2-1400\mu\nu^6|\ .
\end{eqnarray}

There are several patterns here. It is immediately apparent that the   leading logarithm term in the free energy, appearing at torus level, is always:
\begin{equation}
  \labell{eq:freelog}
  F=-\frac{1}{2}\left(\Gamma^2-\frac14\right)
\ln|\mu|=-\frac{m(m+1)}{2}\ln|\mu|\ ,
\end{equation}
when $\Gamma=\frac12+m$ for integer $m$. This gives a leading
contribution to the partition function of the form:
\begin{equation}
  \labell{eq:partipower}
  Z_\Gamma=|\mu|^{\frac12\left(\Gamma^2-\frac14\right)}
=|\mu|^{\frac{m(m+1)}{2}}\ .
\end{equation}

Succinctly, the pattern that emerges is that the partition function
when $\Gamma=\frac12+m$ (for $m$ integer) is the $m$th
Yablonski--Vorobiev polynomial\cite{noumi} (which, interestingly, are related to  Shur functions). The formal zero
energy ground state wave functions of the associated
problem~\reef{eq:eigenvalue} turn out to be ratios of successive
partition functions:
\begin{equation}
  \labell{eq:wavefunctionssimple}
  \psi_\Gamma=e^{-1/\nu\int vdz}=\frac{Z_\Gamma}{Z_{\Gamma-1}}\ .
\end{equation}

\subsection{R--R Flux: Relating The Old Choices To The New}¥
There is another observation to be made. An examination of the
coefficients of the expansion of the free energy $F$ in each case
shows that  the expansion coincides with the $\mu\to-\infty$
expansion that in the integer $\Gamma$ case was associated with R--R
flux!
\begin{eqnarray}
   u&=& \left(\Gamma^2-\frac14\right)\times\nonumber\\
&&\,\, \biggl\{\frac{\nu^2}{z^2}+\left(\Gamma^2-\frac94\right)\times\nonumber\\
&&\hskip1.5cm\biggl[2\frac{\nu^4}{z^5}+7\left(\Gamma^2-\frac{21}{4}\right)\frac{\nu^6}{z^8}+10\left(\Gamma^2-\frac{29}{4}\right)\left(3\Gamma^2-\frac{83}{4}\right)\frac{\nu^8}{z^{11}}\nonumber\\
&&\hskip1.8cm+\frac{13}{64}\left(704\Gamma^6-19216\Gamma^4+178436\Gamma^2-536219\right) \frac{\nu^{10}}{z^{14}}\nonumber\\
&&\hskip1.8cm+\frac{1}{32}\left(23296\Gamma^8-1050880\Gamma^6+18466208\Gamma^4-142989520\Gamma^2+393367971\right)\frac{\nu^{12}}{z^{17}}\nonumber\\
&&\hskip1.8cm+\cdots\biggr]\biggr\}\nonumber\\
F&=& \frac12\left(\Gamma^2-\frac14\right)\times\nonumber\\
&&\,\, \biggl\{-g_s^0\ln|\mu| +\left(\Gamma^2-\frac94\right)\times\nonumber\\
&&\hskip1.5cm\biggl[\frac16g_s^2+\frac16\left(\Gamma^2-\frac{21}{4}\right)g_s^4+\frac19\left(\Gamma^2-\frac{29}{4}\right)\left(3\Gamma^2-\frac{83}{4}\right)g_s^6\nonumber\\
&&\hskip1.8cm+\frac{1}{768}\left(704\Gamma^6-19216\Gamma^4+178436\Gamma^2-536219\right) g_s^8\nonumber\\
&&\hskip1.8cm+\frac{1}{7680}\left(23296\Gamma^8-1050880\Gamma^6+18466208\Gamma^4-142989520\Gamma^2+393367971\right)g_s^{10}\nonumber\\
&&\hskip1.8cm+\cdots\biggr]\biggr\}\ .\labell{eq:negativemuexpand}
\end{eqnarray}

In other words, our solutions correspond to an alternative
non--perturbative completion of the $\mu\to-\infty$ series which, for
$\Gamma$ half--integer, re--sums to a rational function for $u(z)$ and
the logarithm of a Yablonski--Vorobiev polynomial for $F$.

\begin{figure}[ht]
\begin{center}
\includegraphics[scale=0.55]{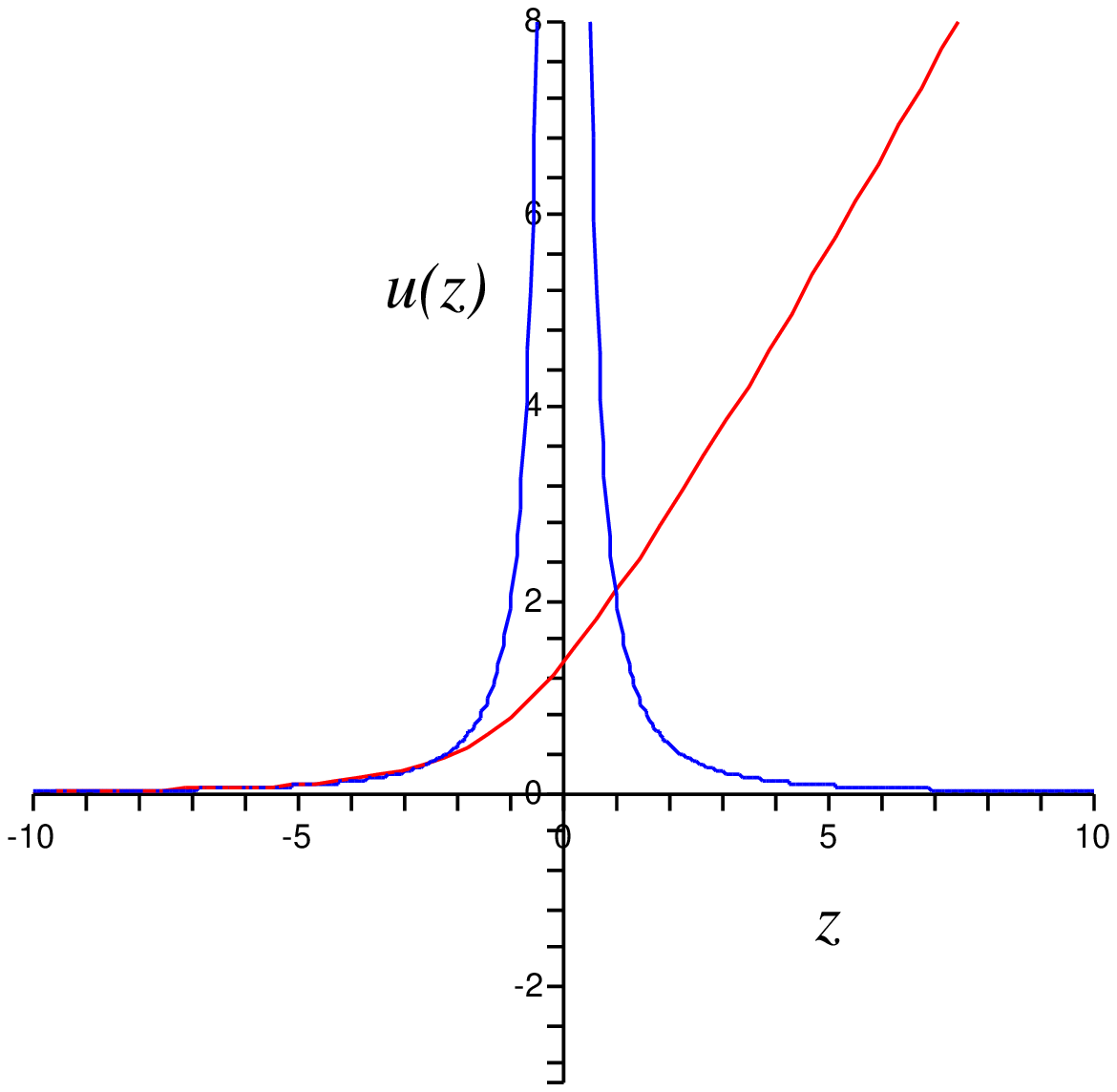}  
\includegraphics[scale=0.55]{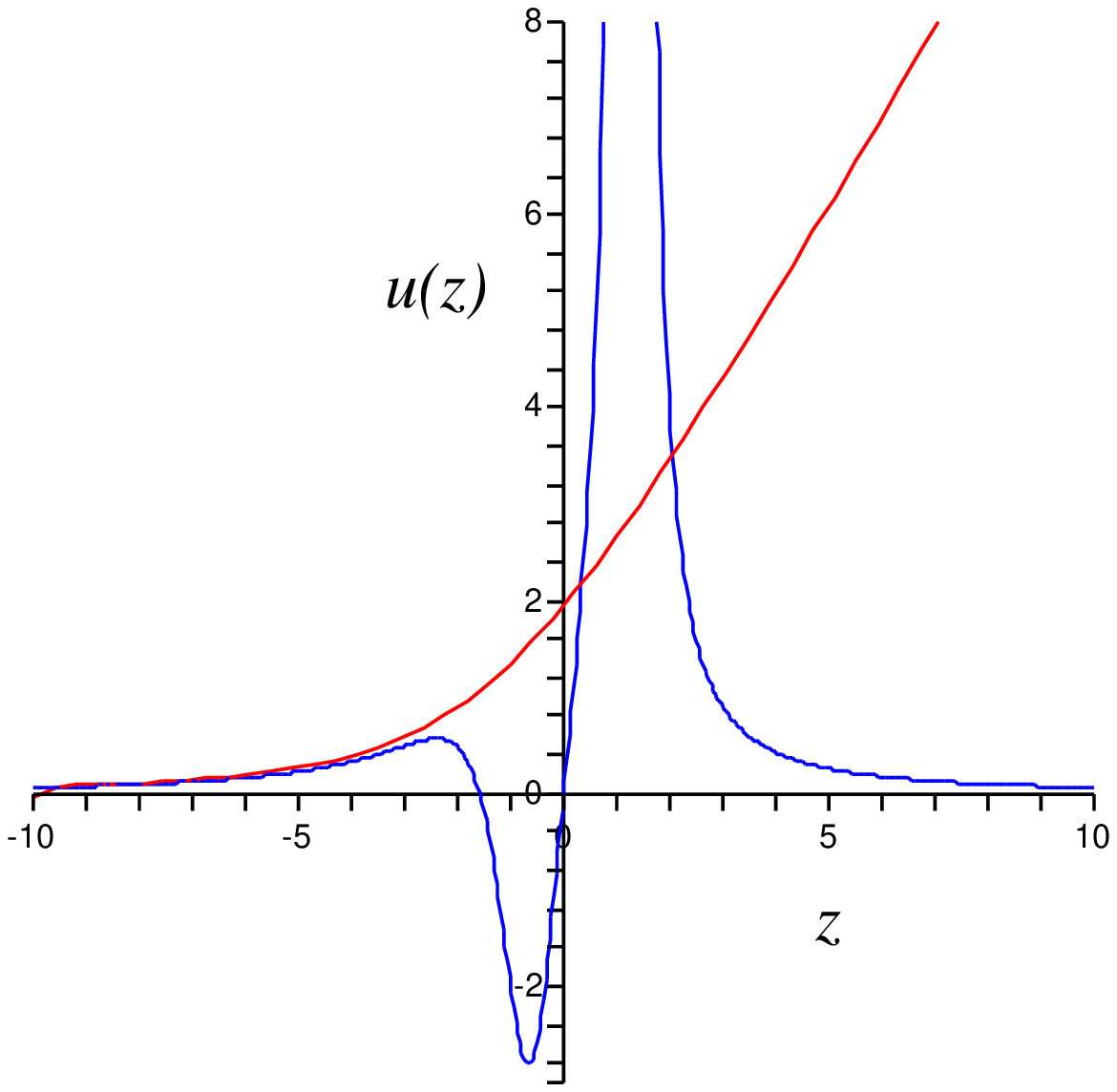}  
\end{center}
\caption{\small Comparison of the $k=1$ solutions   using the standard boundary conditions with the new, rational solutions, for $\Gamma=\frac32$ and $\Gamma=\frac52$, (see equations~\reef{eq:simplecasea} and equation~\reef{eq:nextcase52}).} 
\label{fig:comparison}
\end{figure}

This re--summation gives an exact solution that has the same expansion
for both $\mu\to+\infty$ and $\mu\to-\infty$, and shares the
$\mu\to-\infty$ regime with the previous solutions.  In
figure~\ref{fig:comparison}, we plot examples of $\Gamma=\frac32$ and
$\Gamma=\frac52$, for $k=1$, superimposing the two types of solutions.
 
\subsection{A Family of Special Solutions}

It is worth noting that the $k=1$ case of $\Gamma=\frac32$, which has
$u(z)={2\nu^2}/{z^2}$, is one of an infinite family of exact solutions
of this simple inverse squared form. Every $k$ has such solutions. It
was already noted in ref.\cite{Johnson:2004ut}, that for $k=m$, the
value $\Gamma=\frac12+m$ yields an exact solution
$u(z)=m(m+1)\nu^2/z^2$.  This yields exact expressions for the free
energy and partition function which are given by:
\begin{eqnarray}
  \labell{eq:freelogtwo}
  F&=&-\frac{1}{2}\left(\Gamma^2-\frac14\right)
\ln|\mu|=-\frac{m(m+1)}{2}\ln|\mu|\ ,\nonumber\\
  Z_\Gamma&=&|\mu|^{\frac12\left(\Gamma^2-\frac14\right)}
=|\mu|^{\frac{m(m+1)}{2}}\ .
\end{eqnarray}
In fact, there are even more solutions of this form. If we index
solutions by the integers $(k,m)$, meaning the $k$th model with
$\Gamma=\frac12+m$, and were to make a table of the solutions, with
$m$ increasing downwards indexing rows and $k$ increasing to the right
indexing columns, then the cases referred to above lie on the leading
diagonal. See table~\ref{tab:structure} on
page~\pageref{tab:structure}.  Actually, the entire upper right
triangle above the diagonal also contains solutions of the form, as we
shall now prove.

The equation we are solving is equation~\reef{eq:nonpert} and we set
$t_k$ so that ${\cal R}= cR_k-z$, where $c$ is a constant chosen such
that the coefficient of $u^k$ is unity. The exception is the case
$k=0$, which still has meaning here. In this case, $c$ is actually
zero, and so ${\cal R}=-z$. There are nevertheless solutions for
$u(z)$ since the string equation is now:
\begin{equation}
uz^2+\nu^2\frac14=\nu^2\Gamma^2\ ,
\end{equation}
and so we see that there is a solution
$u(z)=\nu^2\left(\Gamma^2-\frac14\right)/z^2=m(m+1)\nu^2/z^2$ for
every $m$ at $k=0$. This is the first column in our table. The other
special case of interest is the first row, $m=0$, \ie,
$\Gamma=\frac12$. For this case, the $u$--independent terms in the
string equation cancel exactly, and the remaining terms vanish
individually for $u=0$.

To show the existence of several more solutions of the form
$u(z)=A/z^2$, we can proceed by differentiating the string equation
once with respect to $z$.  After doing this, a common factor of~${\cal
  R}$ can be discarded, leaving a very simple
equation\cite{Dalley:1992vr}:
\begin{equation}
u^\prime{\cal R}+2u{\cal R}^\prime-\frac12{\cal R}^{\prime\prime\prime}=0\ ,
\end{equation}
and substituting the form ${\cal R}= cR_k-z$, the recursion
relation~\reef{eq:recursion} for the $R_k$ can be used to show that
\begin{equation}
2cR_{k+1}^\prime+zu^\prime+2u=0\ .
\end{equation}
The key point is that the last two terms cancel each other if
$u(z)=A/z^2$ for $A$ constant. So this leaves us to determine whether
$R_{k+1}^\prime$ can vanish. It obviously does so if $u=0$, recovering
our $m=0$ family above, but there is an additional striking fact
(which we shall again put to great use in later sections): For $k>m$,
the differential polynomials $R_k$ and their derivatives vanish
identically for this form of $u$ when $A=m(m+1)$.  Therefore, the
entire upper right triangle of solutions in table~\ref{tab:structure}
(page~\pageref{tab:structure}) is of the exact form
$u=m(m+1)\nu^2/z^2$.

For the solutions below the diagonal (and away from the $k=0$ column
to the left), more intricate rational forms appear, and can be
generated recursively by use of the explicit B\"acklund
transformations of equation~\reef{eq:Back-Explicit-Gam}, as we did
earlier for $k=1$. For further examples, for $k=2$ we have displayed
the case of $\Gamma=\frac72$ in table~\ref{tab:structure}, and for
$\Gamma=\frac92$ we have:
\begin{equation}
u=\frac{20\nu^2z^3(z^{15}+168z^{10}\nu^4+26208z^5\nu^8-451584\nu^{12})}{(z^{10}-168z^5\nu^4-1344\nu^8)^2}
\labell{km24}
\end{equation}
and at $\Gamma=\frac92$, for $k=3$:
\begin{equation}
u=\frac{20\nu^2(z^{14}+2592z^7\nu^6+155520\nu^{12})}{z^2(z^7-720\nu^6)^2}
\labell{km34}
\end{equation}

We list several entries in table~\ref{tab:structure}, showing the
overall structure of the solution space that we have uncovered. While
the $k=1$ case maps to the Painlev\'e~II equation, for which rational
solutions have been uncovered in the classic work of
ref.\cite{airault}, the other higher $k$ cases represent a large
family of interesting generalisations. Note also the fact that we have
uncovered an infinite family of generalisations of the
Yablonski--Vorobiev polynomials. For example, for the $k=2$, $m=3$
case, the partition function is:
\begin{equation}
  \label{eq:partfunk2m3}
  Z=|\mu^5-24\nu^4|\ ,
\end{equation}
while for $k=2$, $m=4$, it is 
\begin{equation}
  \label{eq:partfunk2m4}
  Z=|\mu^{10}-168\mu^5\nu^4-1344\nu^8|\ .
\end{equation}
These should be compared to their $k=1$ counterparts for $m=2$ and
$m=3$ in equations~\reef{eq:freeparti52} and~\reef{eq:freeparti72}.

\begin{table}[htdp]
\caption{The structure of the large family of rational solutions for $u(z)$ possessed by the string equation~\reef{eq:nonpert}, for $\Gamma=\frac12+m$.}
\label{tab:structure}
\begin{center}
\begin{tabular}{|c|cccccc|}
\hline
\phantom{$k=0$}&$k=0$&$k=1$&$k=2$&$k=3$&$k=4$&$\hdots$\\
\hline
$m=0$&$0$&$0$&$0$&$0$&$0$&$\hdots$\\
$m=1$&$\frac{2\nu^2}{z^2}$&$\frac{2\nu^2}{z^2}$&$\frac{2\nu^2}{z^2}$&$\frac{2\nu^2}{z^2}$&$\frac{2\nu^2}{z^2}$&$\hdots$\\
$m=2$&$\frac{6\nu^2}{z^2}$&$\frac{6\nu^2z(z^3+4\nu^2)}{(z^3-2\nu^2)^2}$&$\frac{6\nu^2}{z^2}$&$\frac{6\nu^2}{z^2}$&$\frac{6\nu^2}{z^2}$&$\hdots$\\
$m=3$&$\frac{12\nu^2}{z^2}$&$\frac{12z\nu^2(z^9+150z^3\nu^4-200\nu^6)}{(z^6-10z^3\nu^2-20\nu^4)^2}$&$\frac{12\nu^2(z^{10}+72z^5\nu^4+96\nu^8)}{z^2(z^5-24\nu^4)^2}$&$\frac{12\nu^2}{z^2}$&$\frac{12\nu^2}{z^2}$&$\hdots$\\
$m=4$&$\frac{20\nu^2}{z^2}$&$\mbox{equation}~\reef{eq:uv92}$&\mbox{equation}~\reef{km24}&\mbox{equation}~\reef{km34}&$\frac{20\nu^2}{z^2}$&$\hdots$\\
$\vdots$&$\vdots$&$\vdots$&$\vdots$&$\vdots$&$\vdots$&$\ddots$\\
\hline
\end{tabular}
\end{center}
\label{default}
\end{table}%

\section{Some Spectra}
\label{sec:spectrum}
As already mentioned, there is an important associated eigenvalue
problem, written in equation~\reef{eq:eigenvalue}.  A solution at
$\lambda=0$ was written in equation~\reef{eq:wavefunctionssimple}, but
the nature of the solutions away from $\lambda=0$ is of interest.

It is useful to write the problem in different
variables~\cite{Carlisle:2005wa}, defining the function $\phi(z)$ and
the variable $x$ by $\psi=x^\frac12\phi(x)$, $x=\lambda^\frac12
z/\nu$.  For the potential $u$, the problem becomes:
\begin{equation}
  \labell{eq:newproblem}
  x^2\frac{\partial^2\phi}{\partial x^2}+x\frac{\partial\phi}{\partial x}+\phi\left(x^2-\frac14-\frac{x^2 u}{\lambda}\right)=0\ ,
\end{equation}
which takes a rather simple form in the large $z$ (or $x$) limit. In
this limit, $u\sim\frac{\lambda}{x^2}\left(\Gamma^2-\frac14\right)$
for all solutions. Our equation then becomes Bessel's equation:
\begin{equation}
  \labell{eq:bessel}
  x^2\frac{\partial^2\phi}{\partial x^2}+x\frac{\partial\phi}{\partial x}+\left(x^2-\Gamma^2\right)\phi=0\ .
\end{equation}
In general, our rational solutions for $u(z)$ define an interesting
family of generalizations of Bessel's functions (remaining asymptotic to them at large $z$ or $x$) that would be
interesting to study. We will leave that for future investigations.

Recalling that there is an infinite family of special cases where
$u=\frac{\lambda}{x^2}\left(\Gamma^2-\frac14\right)$ exactly, it is especially
interesting to study this  exact Bessel function case. In fact,
the relevant solutions are $\phi(x)=J_\Gamma(x)$, where:
\begin{eqnarray}
  \labell{eq:bessels}
  J_n(x)&=&\sum_{l=0}^\infty \frac{(-1)^l}{2^{2l+|n|}l!(|n|+l)!} x^{2l+|n|}\ ,\quad |n|\neq\frac12\ ,\nonumber\\
J_{-\frac12}(x)&=&\sqrt{\frac{2}{\pi x}}\cos x\ ,\nonumber\\
J_{\frac12}(x)&=&\sqrt{\frac{2}{\pi x}}\sin x\ ,
\end{eqnarray}
and this translates, in our original problem, to
$\psi(z)=\nu^{-1/2}\lambda^{1/4}z^{1/2}J_\Gamma(\lambda^{1/2}z/\nu)$.
This is exact for $\Gamma=\frac12+m$ for $k\geq m$, for which the
potential is $u=\lambda m(m+1)/x^2$. The very simplest case is
$\Gamma=\frac12$, for which the potential vanishes. Indeed, we see
that the solutions are just purely sinusoidal or cosinusoidal,
representing free fields.

\section{On Microscopic and Macroscopic Loops}

We have already seen, by examining the perturbative expansion, that
our models don't seem to contain branes. Let us explore this further.
A very useful diagnostic tool in this context is the loop operator
$w(\lambda,\mu)$, and its Laplace transform $w(\ell,\mu)$, which are
related to ${\hat R}(z,\lambda)$, the diagonal of the resolvent of the
Hamiltonian ${\cal H}=-\nu^2\partial_z^2+u(z)$:
\begin{equation}
{\hat R}(z,\lambda)\equiv \langle z| \frac{1}{{\cal H}+\lambda}|z\rangle\ ,
\end{equation}
as follows\cite{Gross:1990vs,Gross:1990aw}:
\begin{equation}
\langle w(\lambda,\mu)\rangle =\int^\mu {\hat R}(z,\lambda) dz=\int_0^\ell \langle w(\ell,\mu)\rangle e^{-\lambda\ell} d\ell\ .
\end{equation}
(Note here that the $\lambda$ in the equations immediately above has
the opposite sign to the $\lambda$ of the previous section. This is a
matter of convention, and removes numerous factors of $(-1)^{1/2}$ in
what follows.) The expectation value $\langle w(\lambda,\mu)\rangle$,
once integrated once with respect to $\lambda$ and divided by $\nu$,
is ---when the familiar choices of section~\ref{sec:oldchoices} are
made--- the free energy of a D--brane probe that lies along the
Liouville direction, the FZZT
D--brane\cite{Fateev:2000ik,Teschner:2000md}. It is of course closely
related to the set of eigenfunctions and eigenvalues
$\{\psi_\lambda(z),\lambda\}$ of~${\cal H}$, which, taken together as
defining a function $\psi(\mu,\lambda)=\psi(z,\lambda)|_{z=\mu}$
of~$\lambda$, may be thought of as the FZZT partition
function\cite{Klebanov:2003wg}. This free energy is also the effective
potential $V_{\rm eff}(\lambda,\mu)$ for one scaled eigenvalue in the
original matrix model language and gives information about the ZZ
D--branes in the theory, as we shall discuss near the end of the next
two subsections~\ref{sec:standardloops} and~\ref{sec:newloops}.

To proceed, we note that the resolvent satisfies the non--linear
equation of Gel'fand and Dikii\cite{Gelfand:1975rn}:
\begin{equation}
4[u(z)+\lambda]{\hat R}^2-({\hat R}^\prime)^2+2{\hat R}{\hat R}^{\prime\prime}=1\  .
\labell{eq:gelfanddikii}
\end{equation}
(Recall that a prime denotes $\nu\partial/\partial z$.)  Calculating
$R(z,\lambda)$ for an arbitrary potential $u(z)$ is a tall order in
general, but often, progress can be made {\it via} the
expansion\cite{Gelfand:1975rn}:
\begin{equation}
{\hat R}(z,\lambda)=\sum_{m=0}^\infty \frac{R_m[u]}{\lambda^{m+\frac12}}\ ,
\labell{eq:expandresolve}
\end{equation}
where the $R_m[u]$, differential polynomials in $u(z)$ and its
derivatives, were defined above in equations~\reef{eq:recursion}
and~\reef{eq:firstfew}.

One might also make sensible progress by developing ${\hat
  R}(z,\lambda)$ as a perturbative expansion in small $\nu$,
corresponding essentially to the string loop expansion. From this, one
can learn about the physics of loops at least order by order in string
perturbation theory.

\subsection{Standard Loops}
\label{sec:standardloops}
For orientation, it might be helpful to first discuss the case where
we have the familiar loop behaviour, in order to understand what is to
come. For the case of $k=1$, with $\Gamma$ integer, we have from the
large positive $z$ expansion, the leading order (sphere level)
solution:
\begin{equation}
u(z)=z+O(\nu)\ .
\labell{eq:spheretraditional}
\end{equation}
We can see directly from the Gel'fand--Dikii
equation~\reef{eq:gelfanddikii} that only the first term on the left
hand side survives in this limit, and pure algebra yields the result:
\begin{equation}
{\hat R}(z,\lambda)=\frac{1}{2(z+\lambda)^{\frac12}}\ .
\labell{eq:resolvedsimple}
\end{equation}
Before proceeding, however, let us arrive at this by the alternative
route afforded by equation~\reef{eq:expandresolve}. The
Gel'fand--Dikii differential polynomials~\reef{eq:firstfew} for the
solution~\reef{eq:spheretraditional} become:
\begin{equation}
R_0=\frac12\ , \quad R_1=-\frac{z}{4}\  , \quad R_2=\frac{3}{16}z^2\  , \quad R_3=-\frac{10}{64}z^3\  , \hdots\ ,
\end{equation}
(neglecting terms of order $\nu$ and higher) and so
\begin{equation}
{\hat R}(z,\lambda)=\frac{1}{2\lambda^{\frac12}}-\frac{z}{4\lambda^{\frac32}}+\frac{3}{16}\frac{z^2}{\lambda^{\frac52}}-\frac{10}{64}\frac{z^3}{\lambda^{\frac72}}+\cdots\ , 
\end{equation}
which, upon inspection, can be seen to be re--summed to give our
expression in equation~\reef{eq:resolvedsimple}.  From the resolvent,
we can evaluate the expectation value of the loop operator:
\begin{equation}
\langle w(\lambda,\mu)\rangle=(\mu+\lambda)^\frac12\ .
\labell{eq:looptraditional}
\end{equation}
There are two interesting perturbative regimes, $\lambda\gg\mu$ and
$\lambda\ll\mu$. The former (the integral of what was seen above) is
an expansion in half integer powers of $\lambda$:
\begin{equation}
\langle w(\lambda,\mu)\rangle =\lambda^{\frac12}+\frac12\frac{\mu}{\lambda^\frac12}-\frac18\frac{\mu^2}{\lambda^\frac32}+\frac{1}{16}\frac{\mu^3}{\lambda^\frac52}-\frac{5}{128}\frac{\mu^4}{\lambda^\frac72}+\cdots
\end{equation}
while the other is in terms of integer powers of $\lambda$:
\begin{equation}
\langle w(\lambda,\mu)\rangle=\mu^{\frac12}+\frac12\frac{\lambda}{\mu^\frac12}-\frac18\frac{\lambda^2}{\mu^\frac32}+\frac{1}{16}\frac{\lambda^3}{\mu^\frac52}-\frac{5}{128}\frac{\lambda^4}{\mu^\frac72}+\cdots
\end{equation}
Now, since $\lambda$ is conjugate to loop length $\ell$ in the action
of the theory, the small $\lambda$ regime should be dominated by the
physics of long loop length $\ell$, and we should expect to see (if
such physics is present) the exponential suppression of the amplitude
$w(\ell,\mu)$ for large $\ell$, in this case $e^{-\mu\ell}$. This is
clear from the full expression~\reef{eq:looptraditional}. On the other
hand, the large $\lambda$ regime will be dominated by the physics of
microscopic loop length $\ell$. In this regime, we find for
$w(\ell,\mu)$:
\begin{equation}
\langle w(\ell,\mu)\rangle=\sum_{k=0}^\infty \frac{\ell^{k+\frac12}}{\Gamma\left(k+\frac12\right)}\langle {\cal O}_k\rangle \ ,\quad \langle {\cal O}_0\rangle =-\frac{\mu^2}{8}\ , \langle {\cal O}_1\rangle =\frac{\mu^3}{16}\ , \langle {\cal O}_2\rangle =-\frac{5\mu^4}{128}\ ,\hdots
\end{equation}
where here, the $\Gamma$ refers to the $\Gamma$--function, not the
parameter $\Gamma$.  It is nice to check that this fits with some of
the other things we have established about the theory's operator
content. The basic object, $u(z)$ is obtained from the free energy by
differentiating twice with respect to~$\mu$ (and setting $\mu=z$), and
so is the two point function $\langle PP\rangle $ of the object that
couples to $\mu$. The operator is sometimes called $P$, the puncture
operator (at least when $k=1,2$), after the operator which measures the
area of worldsheets.  So we have $\langle PP\rangle =\mu$. On the
other hand, we have $\langle {\cal O}_0\rangle =-\frac{\mu^2}{8}$, and
so $\langle P{\cal O}_0\rangle =-\mu/4$. But since $t_0=-4\mu$, we
have $\langle P\dots\rangle =-\frac14\langle {\cal O}_0\cdots\rangle
$, and so we recover that $u=\langle PP\rangle =\mu$. For the operator
${\cal O}_1$, we have the KdV flows of equation~\reef{eq:kdvflows}. An
insertion of ${\cal O}_1$ is differentiation with respect to $t_1$. We
can do this in the presence to two insertions of $P$, in which case
the KdV flows tell us that since $R_2=3\mu^2/16$, we have $\langle
PP{\cal O}_1\rangle =\frac38\mu$. Integrating twice with respect to
$\mu$ we confirm that $\langle {\cal O}_1\rangle =\frac{\mu^3}{16}$.
This can be checked for $\langle {\cal O}_2\rangle $ and so on.

In this way, we see that the microscopic loops that arise in the large
$\lambda$ regime are nothing but the insertions of the point--like
operators corresponding to the closed string sector, as is well
known\cite{Banks:1990df}. We can write the expectation value of the
operators as:
\begin{equation}
<{\cal O}_k>=\int^\mu \!\! R_{k+1}[u(z)] \,\,dz\ .
\end{equation}

Another useful piece of physics that can be extracted from this probe
is the presence and action of point--like D--branes in the theory. The
loop expectation value in equation~\reef{eq:looptraditional} can be
integrated once with respect to $\lambda$, and after dividing by $\nu$
the result gives the effective potential $V_{\rm eff}(\lambda,\mu)$
for one scaled eigenvalue in the original matrix model language. The
potential's extrema (where $\langle w(\lambda,\mu)\rangle$ vanishes)
correspond to the presence of ZZ D--branes, as seen by our probe. The
value of the potential at the extremum gives the action/tension of the
brane\cite{McGreevy:2003kb,McGreevy:2003ep}. In the current example,
we can see that $\langle w(\lambda,\mu)\rangle$ vanishes at finite
$\lambda$, and the resulting brane tension is proportional to
$\mu^{\frac32}/\nu$, which from equation~\reef{eq:stringcoupling}
tells us that the tension is of order $g_s^{-1}$, as it should be for
a D--brane.

Let us now turn to the models in question. We need to compute the
resolvent in this case, using much more complicated expressions for
the function $u(z)$. Given what has gone before, one might hope that
for these rational solutions, some special circumstance might
transpire that may well render the problem more tractable. One might
expect, for example, that once again the expansion of
equation~\reef{eq:expandresolve} in inverse powers of
$\lambda^\frac12$ becomes re--summable into a closed form solution, as
happened for the example above. In fact, the result is even prettier
than that, as we shall see.

\subsection{New Loops: The Special Solutions}
\label{sec:newloops}
Let us start with the simplest solutions, the special family of exact
solutions identified earlier,
\begin{equation}
u(z)=\frac{\nu^2\left(\Gamma^2-\frac14\right)}{z^2}=\frac{m(m+1)\nu^2}{z^2}\ , \quad k\geq m,\quad  \Gamma=\frac12+m\ ,
\end{equation}
In fact, for these, there is a spectacular simplification. For this
form, the Gel'fand--Dikii differential polynomials $R_k[u]$,
equation~\reef{eq:firstfew}, actually {\it vanish identically} for
$k>m$. For example, for $\Gamma=\frac12$ we have that $u=0$, and
therefore the only non--vanishing differential polynomial is the
trivial one, $R_0=\frac12$.  For $\Gamma=\frac32$, we have
\begin{equation}
u=\frac{2\nu^2}{z^2}\ , \quad R_0=\frac12\ , \quad R_1=-\frac12\frac{\nu^2}{z^2}\ ,
\end{equation}
and the other $R_k[u]$ identically vanish.  For $\Gamma=\frac52$, we have
\begin{equation}
u=\frac{6\nu^2}{z^2}\ , \quad R_0=\frac12\  , \quad R_1=-\frac{3\nu^2}{2z^2}\ , \quad R_2=\frac{9\nu^4}{2z^4}\ ,
\end{equation}
with all other $R_k[u]$ vanishing, and finally for $\Gamma=\frac72$, we have 
\begin{equation}
u=\frac{12\nu^2}{z^2}\ , \quad R_0=\frac12\ , \quad R_1=-\frac{3\nu^2}{z^2}\ , \quad R_2=\frac{45\nu^4}{2z^4}\ , \quad R_3=-\frac{225\nu^6}{2z^6}\ ,
\end{equation}
with all other $R_k[u]$ vanishing. This means that in the case of the
special solutions, the expression for the resolvent is exact, starting
at order $\lambda^{-1/2}$ and ending at order $\lambda^{-\Gamma}$,
with precisely $\Gamma+\frac12=m+1$ terms, for example:
\begin{eqnarray}
\Gamma&=&\frac12\  ,\quad {\hat R}(z,\lambda)=\frac{1}{2\lambda^\frac12}\nonumber\\
\Gamma&=&\frac32\ , \quad {\hat R}(z,\lambda)=\frac{1}{2\lambda^\frac12}\left(1-\frac{\nu^2}{z^2\lambda}\right)\nonumber\\
\Gamma&=&\frac52\ , \quad {\hat R}(z,\lambda)=\frac{1}{2\lambda^\frac12}\left(1-\frac{3\nu^2}{z^2\lambda}+\frac{9\nu^4}{z^4\lambda^2}\right)\nonumber\\
\Gamma&=&\frac72\ ,\quad {\hat R}(z,\lambda)=\frac{1}{2\lambda^\frac12}\left(1-\frac{6\nu^2}{z^2\lambda}+\frac{45\nu^4}{z^4\lambda^2}-\frac{225\nu^6}{z^6\lambda^3}\right)\ .
\end{eqnarray}

Now we can integrate once with respect to $z$ to get the loop operator $w(\lambda,\mu)$. For example:
\begin{eqnarray}
\Gamma&=&\frac12\ , \quad \langle w(\lambda,\mu)\rangle =\frac{\mu}{2\lambda^\frac12}\nonumber\\
\Gamma&=&\frac32\ , \quad \langle w(\lambda,\mu)\rangle=\frac{\mu}{2\lambda^\frac12}\left(1+\frac{\nu^2}{\mu^2\lambda}\right)\nonumber\\
\Gamma&=&\frac52\ , \quad \langle w(\lambda,\mu)\rangle=\frac{\mu}{2\lambda^\frac12}\left(1+\frac{3\nu^2}{\mu^2\lambda}-\frac{3\nu^4}{\mu^4\lambda^2}\right)\nonumber\\
\Gamma&=&\frac72\ , \quad \langle w(\lambda,\mu)\rangle=\frac{\mu}{2\lambda^\frac12}\left(1+\frac{6\nu^2}{\mu^2\lambda}-\frac{15\nu^4}{\mu^4\lambda^2}+\frac{45\nu^6}{\mu^6\lambda^3}\right)\ .
\end{eqnarray}
and here we must remember that these are exact expressions for models
$m=0,1,2,3,$ respectively (recall that the inverse squared potential
is exact for the $k$th model when $\Gamma=\frac12+m$, with $k\geq m$.)

There are a number of remarks to be made here. The first is that we
have a finite number of terms in these {\it exact} $\lambda$
expansions, and also in the corresponding $\ell$ expansions in the
Laplace transforms. So looking at the large $\lambda$ limit, which is
dominated by the physics of small $\ell$ loops, we have only a finite
number of operators ${\cal O}_j$ appearing in the theory, the values
$j=0,\hdots,m-1$ for $\Gamma=\frac12+m$, and so far we have $k\geq m$.
The second remark is that while our loop has a small $\ell$ (large
$\lambda$) expansion of a sort similar to the usual behaviour
discussed above, the large $\ell$ (small $\lambda$) regime is very
different. There is no re--summation of the series (since it is of a
finite number of terms) into an expression that at small $\lambda$
supports the characteristic $e^{-\mu\ell}$ behaviour. In other words,
{\it the theory does not possess the large loop FZZT D--brane physics
  seen before.}
  
The same features are responsible for the absence of point--like ZZ
D--branes from the picture painted by our probe as well. The extrema
of the effective potential $V_{\rm eff}(\lambda,\mu)=\nu^{-1}\int
\langle w(\lambda,\mu)\rangle d\lambda$, given by the zeros of the
loop operator, are located where $\lambda$ goes to infinity. So if
there were ZZ D--branes present, they would be located by our probe as
being in the weak coupling regime in the Liouville direction
$\varphi=-\ln\lambda$, but an evaluation of the effective potential
there gives an infinite result.  Again, an infinite number of terms in
the $\lambda^{-1/2}$ expansion are needed allow the loop to have a
zero at finite $\lambda$, and a corresponding finite effective
potential. Such terms are not available in these models because of the
remarkable truncation of the $R_k[u]$, and so we see that we have no
ZZ D--branes. We will return to this in the conclusion
section~\ref{sec:conclusion}.

\subsection{New Loops: The More General Solutions}
Having seen the remarkable simplifications that occurred in the last
subsection,  for the special solutions $u=m(m+1)\nu^2/z^2$, 
let us turn to the more general rational solutions that
occur for  $\Gamma=\frac12+m$ with 
$k<m$. Focusing on $k=1$ for definiteness, let us look at
$\Gamma=\frac52$. The potential is given in
equation~\reef{eq:nextcase52}, and proceeding to evaluate the
Gel'fand--Dikii polynomials, it is a remarkable surprise to find that
once again they vanish beyond a certain order. This time, it is for
order beyond two. This matches the previous pattern: Recall that for
$\Gamma=\frac12+m$, the $R_k$ for $k>m$ vanished, for the special
solutions. In fact, it happens again here, even for these much more
complicated rational solutions. This is the general pattern.

In this $m=2$ example, we have:
\begin{equation}
R_0=\frac12\ ; \quad R_1=-\frac{3z\nu^2(z^3+4\nu^2)}{2(z^3-2\nu^2)^2}\ ; \quad R_2=\frac{9\nu^4z^2}{2(z^3-2\nu^2)^2}\ ,
\end{equation}
with all other $R_k$ vanishing. The resolvent is then:
\begin{equation}
{\hat R}(z,\lambda)=\frac{1}{2\lambda^\frac12}\left(1-\frac{3z(z^3+4\nu^2)}{(z^3-2\nu^2)^2}\frac{\nu^2}{\lambda}+\frac{9z^2}{(z^3-2\nu^2)^2}\frac{\nu^4}{\lambda^2}\right)\ ,
\end{equation}
with an expression for the loop that is:
\begin{equation}
\langle w(\lambda,\mu) \rangle =\frac{\mu}{\lambda^\frac12}+\frac{3\mu^2}{2(\mu^3-2\nu^2)^2}\frac{\nu^2}{\lambda^\frac32}-\frac{3}{2(\mu^3-2\nu^2)^2}\frac{\nu^4}{\lambda^\frac52}\ ,
\end{equation}
together with its Laplace transform:
\begin{equation}
\langle w(\ell,\mu) \rangle =\frac{1}{\sqrt{\pi}}\left(\frac{\mu}{2\ell^\frac12}+\frac{3\mu^2\nu^2\ell^\frac12}{(\mu^3-2\nu^2)^2}-\frac{2\nu^4\ell^\frac32}{(\mu^3-2\nu^2)^2}\right)\ ,
\end{equation}
showing the appearance of a finite number of point--like (closed
string) operators, as before.
\begin{equation}
\langle {\cal O}_{0}\rangle=\frac{3\mu^{2}\nu^{2}}{(\mu^{3}-2\nu^{2})^{2}}\ ; \quad \langle {\cal O}_{1}\rangle=-\frac{3\nu^{4}}{2(\mu^{3}-2\nu^{2})^{2}}\ .
\end{equation}

\section{Point--Like Operators}
So we've established from the previous section that we have indeed a
family of point--like operators organised by the KdV flows, as is
familiar for the case of $\Gamma$ integer, or in the bosonic string.
 
There is a major difference in how the operator content manifests
itself in this situation, however, and it is worth pausing to
appreciate it. In the standard theories, there is an infinite set of
operators ${\cal O}_k$, for which the KdV flows describe the
insertion:
 \begin{equation}
 \langle {\cal O}_0{\cal O}_0{\cal O}_k\rangle\equiv\frac{\partial u}{\partial t_k}=R_{k+1}^\prime[u]\ .
 \end{equation}
 The Virasoro constraints (loop equations for microscopic loops)
 remove\cite{Dijkgraaf:1990rs} an infinite set of these operators from
 the system, by supplying a set of recursion relations between
 correlation functions of the ${\cal O}_k$. These relations can be
 used to eliminate an infinite set of operators, leaving only a
 physical set: For the $k$th model, only the operators ${\cal
   O}_0\cdots{\cal O}_{k-1}$ survive.
 
 In our case, things are much simpler. For a given model, the $k$th,
 turning on some $\Gamma=\frac12+m$ ($m>k$) has the remarkable result
 of removing the effects of all operators ${\cal O}_j$ for $j>m$. This
 is consistent with the structure of table~\ref{tab:structure} (page
 \pageref{tab:structure}), and follows from our observation that the
 $R_{j+1}$ vanish for $j>m$. This is before the action of the Virasoro
 constraints which presumably reduces the operator content back to
 only ${\cal O}_0\cdots{\cal O}_{k-1}$, for this $k$th model.

In the case of the $k$th model, switching on instead $k$ units (\ie,
setting $m=k$) reduces the model to $u=m(m+1)\nu^2/z^2$, (we are on
the diagonal of table~\ref{tab:structure}) and the operator content is
precisely the minimal set ${\cal O}_0\cdots{\cal O}_{k-1}$ with no
appeal to Virasoro.
  
If instead for the $k$th model we have $m<k$, then we are in the upper
right triangle of table~\ref{tab:structure} and the potential is still
$u=m(m+1)\nu^2/z^2$, and there are at most $m$ operators: ${\cal
  O}_0\cdots{\cal O}_{m-1}$. The extreme case of this is of course
$m=0$. There we see, that regardless of the value of $k$, the model is
controlled by the completely trivial potential $u=0$. There is no
operator content. This will all play a role in our interpretive
discussion in the conclusions section~\ref{sec:conclusion}.

\section{Phase Transitions}
\begin{figure}[ht]
\begin{center}
\includegraphics[scale=0.55]{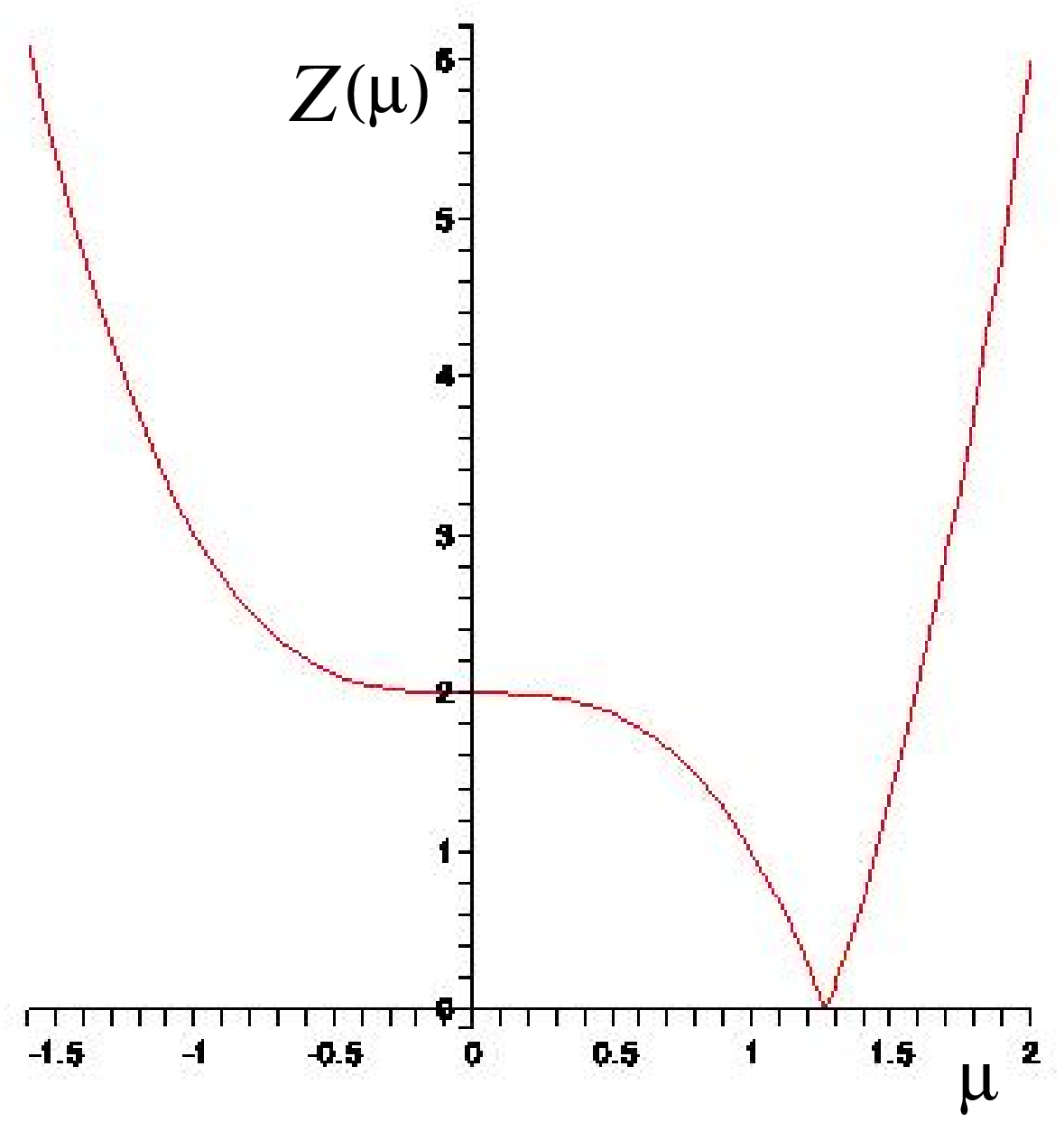}\includegraphics[scale=0.55]{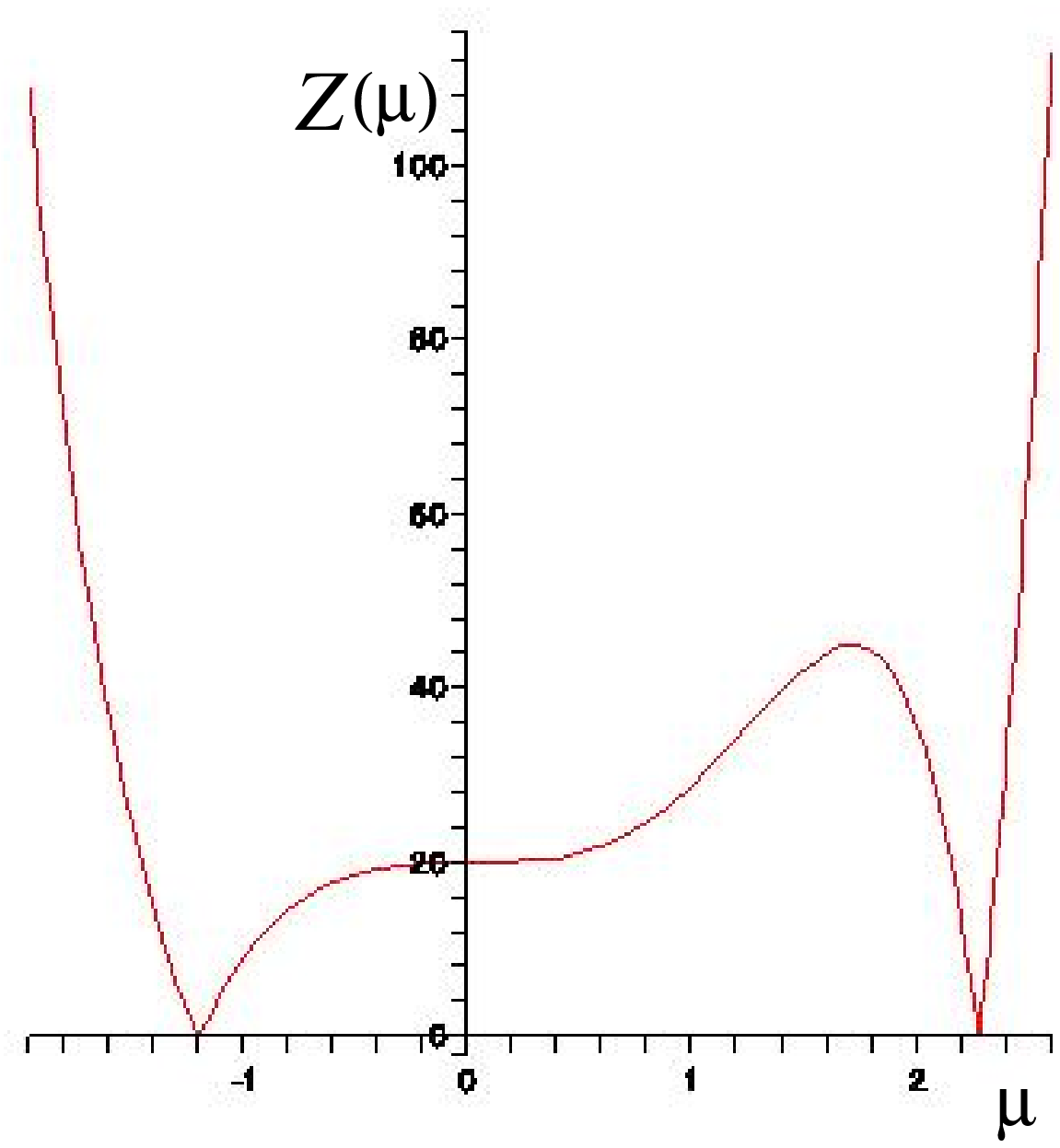}  
\end{center}
\caption{\small The partition function $Z(\mu)$ as a function of $\mu$ for the $k=1$ cases of $\Gamma=\frac52$ and $\frac72$. There is a phase transition whenever there is a zero.} 
\label{fig:transition}
\end{figure}

An unusual feature of our models is the existence of points on the
real $\mu$ line where the partition function vanishes. See
figure~\ref{fig:transition} for $k=1$ examples
$\Gamma=\frac52,\frac72$. At such points, the free energy goes through
a discontinuity, and there are poles in all correlation functions of
point--like operators.

A zero of the partition function has its origins in a double pole of
$u(z)$. The appearance of a double pole in the function $u(z)$ has
been considered before, in the context of the non--perturbative
physics uncovered in the original double scaling limit of
refs.\cite{Brezin:1990rb,Douglas:1990ve,Gross:1990vs}.  The
accompanying physics was originally
conjectured\cite{Brezin:1990rb,Banks:1990df} to be indicative of a
phase transition, but this idea was left undeveloped as the discussion
was overtaken by other more pressing physical issues. Among them were:

\begin{enumerate}
\item The controlling string equation was not our equation~\reef{eq:nonpert}, but instead
  Painlev\'e~I:
\begin{equation}
-\frac{\nu^2}{3} \frac{\partial^{2}u(z)}{\partial z^{2}}+u^{2}=z\ ,
\labell{eq:PainleveI}
\end{equation}
with a large $z$ asymptotic $u\sim z^{1/2}+O(\nu)$. For a real
solution with that asymptotic, there is an infinite number of double
poles on the real axis, and the location of the first such pole in not
fixed by perturbation theory. So there is a non-perturbative parameter
corresponding to the location, $z_{0}$ of the first pole.
 
\item With a pole in $z$, and the above asymptotic, there is
  inevitably only a discrete spectrum for the Hamiltonian~${\cal H}$.
  This discrete spectrum, with eigenvalues $\lambda_{i}$, gives poles
  in $\lambda$ in the loop $\langle w(\lambda)\rangle$, which
  non--perturbatively violate\cite{David:1990ge,David:1991sk} the loop
  equations for $\langle w(\lambda)\rangle$.
\end{enumerate}

In our case, we avoid these issues neatly, since:

\begin{enumerate}
\item The physics supplies us with  a different string equation, given by
  equation~\reef{eq:nonpert}, and different solutions.  The locations
  of the poles of our rational functions are fixed. There are no
  non--perturbative parameters to be determined. Let us denote the
  location of the first pole coming from $z=+\infty$ as $z_{0}$. It is
  a fully determined quantity.
  
\item There is always a continuous spectrum on $(z_{0},+\infty)$,
  given by our studies in section~\ref{sec:spectrum}, the
  wavefunctions being Bessel functions for the special solutions, and
  asymptotically so for the general case. There are therefore no
  $\lambda$--poles in the loop $w(\lambda)$, and from the modern
  perspective, this ensures that the Liouville coordinate (at least as
  seen by this probe), $\varphi=-\ln\lambda$, is continuous. There is
  of course a discrete spectrum that must arise between two successive
  poles, for rational solutions for which there are multiple
  $z$--poles. Perhaps those are disconnected regions upon which we do
  not consider our physics. We need not do so since our wavefunctions
  vanish at the boundaries of these regions and so there can be no
  tunneling. Logically, it is possible that an examination of the loop
  equations for these models may yield consistent physics for those
  regimes as well. There is the intriguing possibility that these
  regimes represent physics where these probes see the Liouville
  direction $\varphi$ as having fractionated into discrete points.
  This is a subject for future study.
\end{enumerate}

So now that we have a very well-behaved system, we are free to
re-examine the meaning of the transitions at the zeros of $Z(\mu)$. At
present, it is not at all clear what the physical interpretation of
the transition is, and whether earlier
speculations\cite{Brezin:1990rb,Banks:1990df} about this being a
``condensation of handles'' are relevant here.  This seems unlikely
here (although there may be surprises), given that one of the
hallmarks of our physics seems to be the very gentle growth of genus
perturbation theory. So we might look elsewhere for an interpretation,
and have not yet found one.

It is interesting to note some general properties of the transition.
Essentially, the properties of the partition function in the
neighbourhood of the zeros is captured by the special solutions
$u=m(m+1)\nu^2/z^2$, and so we can focus on them. Recall that the
partition function is $Z(\mu)=|\mu|^\frac{m(m+1)}{2}$, and so except
for the case of $m=1$, its first $\mu$--derivative at the transition
vanishes.  For $m=1$, it has a jump discontinuity from $-1$ to $+1$.
The quantity $\partial Z/\partial\mu$ may therefore be part of a
definition of a useful order parameter. The natural quantity for all
$m$ might in fact be $\mu(\partial Z/\partial\mu)$ since it vanishes
at the transition point for all $m$. We will leave the study of these
transitions for future work, but expect that the properties above will
play an important role in characterising the physics.

\section{Conclusion}
\label{sec:conclusion}
A useful way to restate some of our key results seems to be as
follows. For integer $\Gamma$ we know that this regime represents
background R--R flux, which can be described in the worldsheet theory
in terms of insertions of a closed string operator into all
correllators, etc. For half integer $\Gamma=\frac12+m$, ($m\in\IZ^+$),
we still have an expansion in terms of closed string world sheets, and
it seems economical to think of $\Gamma$ as representing an aspect of
a background in the closed string theory (although it might be
premature to think of it as simply half--integer R--R flux before
further comparison with an explicit computation from the continuum
formulation). Setting $\Gamma=\frac12$ results in only $u=0$ as the
solution to the string equation. All operators are trivial, and the
theory has no content.  One can therefore think of the $\Gamma=\frac12$
background as simply screening the effects of all operators in the
theory, rendering it trivial.  Switching on $m$ has the interesting
effect that it allows (or unscreens) $m$ operators, ${\cal
  O}_0,\cdots, {\cal O}_{m-1}$, yielding non-trivial physics.

Interestingly, the theory cannot support D--branes however, since the
probe of their presence ---the loop operator $w(\lambda,\mu)$--- needs
an infinite number of the ${\cal O}_j$ operators to be non--zero in
its large $\lambda$ expansion, so that it can be re--summed to
reconstruct the required small $\lambda$ (large loop length $\ell$)
behavior. These are simply not available at finite $m$.  

From this perspective we can see how (at least formally) to return to
the types of string theory we are familiar with. We take a large $m$
limit. In such a limit, where we can indeed fill out the expansion of
the loop operator and allow it to support large loops. The presence of
so many more of these operators presumably will also increase the rate
of growth of string perturbation theory so that there are
$\exp(-C/g_s)$ effects again, signalling the presence of
D--branes\cite{Shenker:1990uf}.  This will make sense at the level of
solutions to the string equation too, and one would expect that the
terms in the expansion of the function $u(z)$ will return to the form
we are familiar with for integer $\Gamma$, (see
equation~\reef{eq:negativemuexpand}), and we can match on to the
regime containing background D--branes.  This would be interesting to
study.

This family of models therefore represents a rather clear example of
the contrast between the types of string theories that most are
familiar with (with asymptotic expansions in small $g_s$, and branes)
and this new type of model which clearly is very simple (re--summable
expansions and no branes) but nonetheless, we submit, an instructive
type of string theory.  We've been able to identify aspects of the
path that connects these, by explicitly following the operator
content.

This may be more than an exercise for its own sake. There may well be
situations involving the types of string theory we wish to use for
other applications (including perhaps the study of Nature) where the
background (an exotic choice of flux, perhaps) induces the type of
behaviour in these models. Such models may therefore furnish an
effective description of some important physical subsector of other
stringy scenarios.

\bigskip
\bigskip

\section*{Acknowledgments}
CVJ wishes to thank Ofer Aharony, Tameem Albash, David Berenstein,
Michael Gutperle, Ramakrishnan Iyer, Rob Myers, Jeffrey Pennington, and
Joe Polchinski for useful comments and conversations. Thanks also to
James Carlisle for an early suggestion to revisit the role of poles in
the susceptibility. This work is supported by the DOE. Some of this
work was carried out at the Centre International de Rencontres
Math\'ematiques (CIRM) in Luminy, during the workshop entitled
``Affine Hecke algebras, the Langlands program, Conformal field theory
and Matrix models''. Thanks to the organizers of the workshop, and the
staff of the centre, for providing a particularly pleasant and
stimulating atmosphere. Some of this work took place at the Aspen
Center for Physics. Thanks to the staff at the centre for an excellent
working environment.

\providecommand{\href}[2]{#2}\begingroup\raggedright\endgroup

\end{document}